\documentclass[onecolumn,floatfix,aps,nofootinbib,showkeys,10pt]{revtex4-2}

\usepackage[dvips]{epsfig}
\usepackage[english]{babel}
\usepackage{bbm}
\usepackage{verbatim}
\usepackage{array}
\usepackage{float}
\usepackage{bm} 
\usepackage{amsmath}
\usepackage{amsfonts}
\usepackage{amssymb}
\usepackage{hyperref}
\usepackage[dvipsnames]{xcolor}
\hypersetup{colorlinks=true,linkbordercolor=Red,linkcolor=Red, citecolor=Red}
\usepackage{indentfirst} %primeiro paragrafo com espaço
\usepackage{float}

\usepackage{bbding}
\usepackage{booktabs}

\usepackage{bbold}

\begin{document}

\title{Revisiting Takahashi's inversion theorem in discrete symmetry-based dual frameworks}

\author{R. J. Bueno Rogerio}
\email{rodolforogerio@gmail.com}
\affiliation{Instituto de F\'isica e Qu\'imica, Universidade Federal de Itajub\'a - IFQ/UNIFEI,
Av. BPS 1303, CEP 37500-903, Itajubá - MG, Brazil.}

\author{R.~T.~Cavalcanti}
\email{rogerio.cavalcanti@unesp.br}
 \affiliation{Departamento de F\'isica, Universidade Estadual Paulista (Unesp), Guaratinguet\'a 12516-410, Brazil}

\author{J.~M.~Hoff da Silva}
\email{julio.hoff@unesp.br}
 \affiliation{Departamento de F\'isica, Universidade Estadual Paulista (Unesp), Guaratinguet\'a 12516-410, Brazil}

\author{C.~H.~Coronado Villalobos}
\email{ccoronado@utp.edu.pe}
\affiliation{Universidad Tecnológica del Perú, Lima-Perú}


\date{\today}

\begin{abstract}
The so-called Takahashi's \emph{Inversion Theorem}, the reconstruction of a given spinor based on its bilinear covariants \cite{Takahashi:1982bb}, are re-examined, considering alternative dual structures. In contrast to the classical results, where the Dirac dual structure plays the central role, new duals are built using the discrete symmetries $\mathcal{C}, \mathcal{P}, \mathcal{T}$. Their combinations are also taken into account. Furthermore, the imposition of a new adjoint structure led us also to re-examine the representation of the Clifford algebra basis elements, uncovering new bilinear structures and a new Fierz aggregate. Those results might help construct theories for new beyond standard model fields.    
\end{abstract}

%\pacs{}
%\keywords{}

\maketitle

\section{Introduction}\label{intro}

The Dirac theory of spin 1/2 particles is unquestionably one of the outstanding achievements of modern physics. However, despite its success and intense investigation over the last century, up to quite recently, a subtle feature of the Dirac theory has not been adequately explored—namely, the spinorial dual structure. The theory's predictions heavily depend on the standard dual structure of the theory, dismissing a large branch of possibilities that could be applied to the description of new beyond the standard model fields. The investigation of mass-dimension-one spinors (Elko) \cite{Ahluwalia:2004ab} approach this shortsightedness by calling the community's attention to the possibility of building, from first principles, a candidate to describe dark matter\footnote{See \cite{Ahluwalia:2019etz,Ahluwalia:2022ttu} for a recent and comprehensive discussion.}. For such, consistency constraints led to a deep and careful mathematical analysis of some fundamental aspects of the Quantum Field Theory by looking toward defining physical fields that carry relevant physical information. Moreover, it has paved the road for a broader research front, from cosmology \cite{Pereira:2017bvq, Pereira:2018hir, Pereira:2018xyl}, to quantum field phenomenology \cite{Lee:2014opa, Lee:2015sqj, Lee:2018ull, Lee:2019fni} and mathematical physics \cite{Fabbri:2010ws,daRocha:2011yr, Cavalcanti:2014uta, Fabbri:2017xyk, Fabbri:2019vut, Fabbri:2020ezx, Rogerio:2022tsl}. A consistent investigation has also been carried out to explore the consequences and generality of the new duals \cite{BuenoRogerio:2017lim, HoffdaSilva:2019ykt, Cavalcanti:2020obq, HoffdaSilva:2022ixq}.

As it is well-known, spinorial physical observables must be represented by real quadratic functionals of a given quantum system. Formally, those observables are represented by the quantities called bilinear covariants. The so-called Lounesto spinor classification \cite{lounestolivro, Cavalcanti:2014wia} stands for an exhaustive categorization based on the bilinear covariants (quadratic functionals), disclosing the possibility of a large variety of spinors. Such a classification encompasses regular and singular spinors, which includes the cases of Dirac, Weyl, and Majorana as a very particular case. After the Lounesto classification, regular spinors can be defined as spinors fields whose both scalar ($\sigma$) and pseudo-scalar ($\omega$) bilinear structures are not both identically null. Singular spinors, on the other hand, have scalar and pseudo-scalar bilinear structures that vanish identically \cite{lounestolivro}. Besides, singular spinors usually are split into three sub-classes, namely flag-pole, flag-dipole, and dipole spinors,  where Majorana and Weyl spinors are the most representative cases. Nonetheless, all the bilinear covariants are bounded to obey certain geometrical conditions, the so-called Fierz-Pauli-Kofink (FPK) identities \cite{lounestolivro,Takahashi:1982wi}.    

A crucial point regarding the Lounesto classification is that it is built upon the usual Dirac dual structure. Hence, if other duals are allowed, one should revisit the Lounesto classification in light of the new duals \cite{CoronadoVillalobos:2020yvr}. The standard Dirac dual is usually presented in textbooks as the most fundamental dual definition for spin one-half fermions. However, as mentioned before, considering alternative fields beyond the standard model might lead to formal inconsistencies solved by a suitable choice of a dual redefinition \cite{Ahluwalia:2004ab,Ahluwalia:2019etz,Ahluwalia:2020miz,HoffdaSilva:2019ykt,Cavalcanti:2020obq}. Upon fixing the appropriate dual definition for a non-standard spinor field, one faces the formal constraints imposed by the so-called Fierz-Pauli-Kofink identities \cite{Crawford:1985qg,Takahashi:1982bb}. In order to preserve such identities after a dual redefinition, following the classical Crawford \cite{Crawford:1985qg} prescription, the underlying Clifford algebra might be deformed \cite{HoffdaSilva:2016ffx}, opening a new branch of formal investigation regarding the internal consistency of the proposed spinor fields and duals. A formal tool for recovering the spinor from its physical information is the so-called inversion theorem proposed by Yasushi Takahashi \cite{Takahashi:1982bb}. It consists of a protocol to reconstruct a given spinor in terms of its bilinear structures. The main mathematical requirement is that such structures are bound to obey the Fierz-Pauli-Kofink identities \cite{Crawford:1985qg,Takahashi:1982bb}. Such a reconstruction is not unique, and the spinor is determined up to a phase factor.

In this paper, we look towards a crevice in Takahashi's formalism by exploiting the freedom given by dual structure redefinition. Given the generality of Takahashi's protocol, we may check and apply it to any well-defined adjoint structures. For that sake, we start by employing a piece of extended machinery, first obtaining a new set of \emph{deformed} bilinear structures satisfying the correspondent algebraic Fierz-Pauli-Kofink identities. Once such a procedure is accomplished, we, thus, are able to check the inversion theorem upon new duals and reconstruct any allowed spinorial structure. The results closely discuss two fundamental aspects of modern spinor fields theory, connecting the formulation of alternative duals and discrete symmetries.

The paper is organized as follows: In the next section, we review the main points on the Lounesto classification, the FPK identities, and Takahashi's inversion theorem. Then, in Sec. \ref{teoremageral}, we exploit the freedom on the adjoint spinorial structure and redefine the inversion theorem in light of the new bilinear structures. Here we are interested in redefining the duals by applying the discrete symmetry operators, $\mathcal{C}, \mathcal{P}, \mathcal{T}$, and their combinations. In Sec. \ref{extension_lounesto}, we investigate the possible new classes, extending the classes unveiled by the new duals in the Lounesto classification. Finally, in Sec. \ref{remarks} we scrutinize the consequences and possible relevance of the new duals built upon the discrete symmetry operators. 

\section{Underlying aspects of the bi-spinor algebra}

Let $M$ denote the Minkowski spacetime. Spinor fields are objects of the spinor bundle  $\mathbf{P}_{\mathrm{Spin}_{1,3}^{e}}(M)\times_{\tau }\mathbb{C}^{4}$, associated to $M$ and carrying $\tau={\left(1/2, 0\right)}\oplus{\left(0, 1/2\right)}$ representations of the spin group  
\cite{Mosna:2002fr, Rodrigues:2002sg, Vaz:2016qyw}.  It is well known that its observable properties are encoded in the so-called bilinear covariants, which are sections of the exterior bundle $\Omega(M)$ \cite{lounestolivro,Crawford:1985qg}. Those symmetry-preserving bilinear covariants come from the multivector structure of the spacetime Clifford algebra. On the standard Dirac theory, they are given by\footnote{In the follwing we shall adopt $\gamma_5:=-i\gamma_{0123}=-i\gamma_{0}\gamma_{1}\gamma_{2}\gamma_{3}$.} 
\begin{subequations}
\label{covariantes}
\begin{align}
\sigma=&\psi^{\dag}\gamma_{0}\psi \in\Omega^0(M),  \\
 \mathbf{J}=&\psi^{\dag}\mathrm{\gamma_{0}}\gamma_{\mu}\psi\;\gamma^{\mu} \in\Omega^1(M),\\
 \mathbf{S}=&\psi^{\dag}\mathrm{\gamma_{0}}\textit{i}\gamma_{\mu\nu}\psi\gamma^{\mu}\wedge\gamma^{\nu} \in\Omega^2(M),\\ 
\mathbf{K}=&\psi^{\dag}\mathrm{\gamma_{0}}\gamma_{\mu}\textit{i}\mathrm{\gamma_{0123}}\psi\;\gamma^{\mu} \in\Omega^3(M),\\
 \omega=&-\psi^{\dag}\gamma_{0}\gamma_{0123}\psi \in\Omega^4(M). 
\end{align} 
 \end{subequations}
The set $\{\mathbbm{1},\gamma_{I}\}$, where $I\in\{\mu, \mu\nu, \mu\nu\rho, {5}\}$ is a composed index, is a basis for ${\cal{M}}(4,\mathbb{C})$ satisfying  the clifford algebra relation $\lbrace\gamma_{\mu }\gamma _{\nu}\rbrace=2\eta_{\mu \nu }\mathbbm{1}$. As usual, $\eta_{\mu\nu}$ denotes the Minkowski spacetime metric tensor. The homogeneous elements of the spacetime Clifford algebra are then chosen accordingly \cite{Crawford:1985qg}  
\begin{equation}\label{cliffordbasis}
\Gamma = \lbrace \mathbbm{1}, \gamma_{\mu}, \gamma_{0123}, \gamma_{\mu}\gamma_{0123}, \gamma_{\mu}\gamma_{\nu}\rbrace. 
\end{equation}
Bearing in mind the usual Dirac adjoint $\bar{\psi} = \psi^{\dag}\gamma_0$, it allows to summarize the bilinear densities relations in \eqref{covariantes} as usual, i. e., $\rho_i = \bar{\psi}\Gamma_i\psi$.

Classically, spinor fields are classified according to irreducible representations of the spin group $\mathrm{Spin}_{1,3}^{e}$. The most well-known classes encompass the Weyl, Dirac, and Majorana spinors, which correspond to different irreducible representations of the spin group. However, Lounesto has shown that another classification can be defined by using bilinear covariants \cite{lounestolivro}. As it is based solely on the multivector structure of the spacetime Clifford algebra, it does not depend on any specific representation. The now well-known Lounesto classification provides a valuable tool for understanding the properties of spinor fields. Notably, we can gain insights into the physical nature of the new spinor field beyond the standard model. Following the usual nomenclature, we call regular spinors those with at least one of the bilinear covariants $\sigma$ or $\omega$ non-null, comprising the classes 1, 2, and 3 on Table \ref{lounesto_classes}. On the other hand, we call singular spinors those with $\sigma=0=\omega$. Thus the classes 4, 5, and 6 on Table \ref{lounesto_classes}. The classes are all defined below:

\begin{table}[H]
    \centering
\begin{tabular}{c|cccccccc}
    Bilinears && Class 1 & Class 2 & Class 3 & Class 4  & Class 5& Class 6& \\ \bottomrule%\midrule
    $\sigma$ && $\neq 0$ & $\neq 0$ & 0 & 0 & 0 & 0 &   \\ 
    $\mathbf{J}$ && $\neq 0$ & $\neq 0$ & $\neq 0$ & $\neq 0$ & $\neq 0$ & $\neq 0$ &   \\ 
    $\mathbf{S}$ && $\neq 0$ & $\neq 0$ & $\neq 0$  & $\neq 0$ & $\neq 0$ & 0 &   \\ 
    $\mathbf{K}$ && $\neq 0$ & $\neq 0$ & $\neq 0$ & $\neq 0$ & 0 & $\neq 0$ &   \\ 
    $\omega$ && $\neq 0$ & 0 & $\neq 0$ & 0 & 0 & 0 &   \\ \bottomrule
\end{tabular}
\caption{Lounesto classification of spinor fields.}
\label{lounesto_classes}
\end{table}
Within the Dirac theory, the above bilinear covariants are usually interpreted as: the mass of the particle ($\sigma$), the pseudo-scalar ($\omega$) relevant for parity-coupling, ($\mathbf{J}$) stands for the current of probability, the direction of the spin ($\mathbf{K}$), and the intrinsic electromagnetic momentum ($\mathbf{S}$), all of them associated to the electron. A fundamental requirement for Lounesto's spinors classification is that all the bilinear covariants must satisfy quadratic algebraic relations, the so-called Fierz-Pauli-Kofink (FPK) identities, which read
\begin{eqnarray}
\label{fpk1}
\mathbf{J}^{2}=\omega^{2}+\sigma^{2},\quad\mathbf{K}^{2}=-\mathbf{J}%
^{2},\quad\mathbf{J}\lrcorner\mathbf{K}=0,\quad\mathbf{J}\wedge\mathbf{K}%
=-(\omega+\sigma\gamma_{0123})\mathbf{S}.  
\end{eqnarray}

Once we have introduced the building blocks of spinor fields and the related bilinear covariants, we can discuss the inversion theorem, which states that once the bilinear covariants are known, the related spinor field can be reconstructed entirely up to a phase factor \cite{Takahashi:1982bb,Takahashi:1982wi,Crawford:1985qg}. In fact, from the bilinear covariants, we can define a multivector object denoted by $\mathbf{Z}$ and known as Fierz aggregate\footnote{Assuming that the bilinear covariants $\sigma,\omega,\mathbf{J},\mathbf{S},\mathbf{K}$ satisfy the Fierz identities.} \cite{Crawford:1985qg,lounestolivro}, 
\begin{eqnarray}
\mathbf{Z}=\sigma + \mathbf{J} +i\mathbf{S}+i\mathbf{K}\gamma_{0123}+\omega \gamma_{0123}. \label{Z}
\end{eqnarray} 
Now, given and arbitrary constant spinor $\xi$ such that $\xi^{\dagger}\gamma_{0}\psi\neq 0$, the spinor $\psi$, related to the bilinear covariants in $\mathbf{Z}$, can be reconstructed by
\begin{equation}
\psi=\frac{1}{4N}\;e^{-i\theta}\mathbf{Z}\xi. \label{3}
\end{equation}
Here $e^{-i\theta}=\frac{1}{N} \xi^{\dagger}\gamma_{0}\psi$ is a phase factor and $N=\frac{1}{2}\sqrt{\xi^{\dagger}\gamma_{0}\mathbf{Z}\xi}$. For the sake of completeness, its adjoint structure yields 
\begin{eqnarray}\label{3dual}
\stackrel{\neg}{\psi} = \frac{1}{4N^*}\xi^{\dag}\mathbf{Z}^{\dag}\gamma_0 e^{i\theta}.
\end{eqnarray}

Established the above reconstruction, we expect that the underlying bilinear covariants could also be algebraically recovered from the Fierz aggregate. In fact, it furnishes
\begin{eqnarray}\label{bilinearreconstruido}
\rho_i=\frac{1}{16|N|^2}\xi^{\dag}\mathbf{Z}^{\dag}\gamma_0\Gamma_i\mathbf{Z}\xi.
\end{eqnarray}
We highlight that for singular spinors, the Fierz identities are, in general, replaced
by the more general conditions \cite{Crawford:1985qg}:
\begin{align}\label{fpk2}
\mathbf{Z}^{2}=4\sigma \mathbf{Z},\nonumber\\ 
\mathbf{Z}\gamma_{\mu}\mathbf{Z}=4J_{\mu}\mathbf{Z},\nonumber\\
 \mathbf{Z}i\gamma_{\mu\nu}\mathbf{Z}=4S_{\mu\nu}\mathbf{Z},\\
\mathbf{Z}i\gamma_{0123}\gamma_{\mu}\mathbf{Z}=4K_{\mu}\mathbf{Z},\nonumber\\
 \mathbf{Z}\gamma_{0123}\mathbf{Z}=-4\omega \mathbf{Z}.\nonumber
\end{align}
The aggregate plays a central role within the Lounesto classification since, in order to complete the classification itself, $\mathbf{Z}$ have to be promoted to a boomerang, satisfying the relation $\mathbf{Z}^{2}=4\sigma\mathbf{Z}$.
This condition is trivially satisfied for the regular spinors case, and $\mathbf{Z}$ is automatically a boomerang. Nonetheless, for singular spinors case, it is not such a straightforward procedure \cite{lounestolivro}.

\section{Re-examination of the Inversion theorem}\label{teoremageral}

As mentioned, the Dirac theory is entirely based on the standard adjoint. However, despite its great success, consistent results of the last decade have shown that different possibilities of duals could play an essential role in spinor field theories beyond the standard model. From the formal side, general features of those duals were investigated from their deep roots in Clifford algebras \cite{HoffdaSilva:2019ykt}, as well as some unexpected underlying algebraic structures \cite{Cavalcanti:2020obq}. From the theoretical side, Elko and mass dimension one spinors are concrete examples of alternative dual requirements \cite{Ahluwalia:2022ttu}. In this context, classical theoretical results of spinor fields must be revisited and re-established. In this section, we shall revisit the results of the previous one in light of the alternative dual spinors built upon discrete symmetries. As shown in Ref. \cite{epldharam}, exciting theoretical possibilities have been raised in the framework of spinor fields whose dual is composed with the aid of operators other than the identity. It could lead to the advent of pseudo-hermitian theories, which, in turn, opens the possibility for statistics different from the usual \cite{lecla,npb4}.  

Following the above discussion, let us start by considering a more general spinorial adjoint structure, which reads
\begin{eqnarray}\label{novodual}
\stackrel{\neg}{\psi} = [\Delta\psi]^{\dag}\gamma_0,
\end{eqnarray}
where the operator $\Delta = \mathcal{P}, \mathcal{C}, \mathcal{T}$ or compositions of such operators, being $\mathcal{P}, \mathcal{C}, \mathcal{T}$ parity, charge-conjugation, and time-reversal symmetry operators, respectively. As usual in the spinorial context, the parity operator is identified by $\mathcal{P} = m^{-1}\gamma^{\mu}p_{\mu}$ \cite{Speranca:2013hqa}. The charge conjugation operator is usually written as $C=\gamma_2 \mathcal{K}$, where the $\mathcal{K}$ operator complex conjugate on the left, and time-reversal operator being $\mathcal{T}=i\gamma_5\gamma_2 \mathcal{K}$ \cite{Ahluwalia:2019etz}. Detailed comments on the $\Delta$ structure and physical/mathematical constraints can be found in \cite{HoffdaSilva:2022ixq,HoffdaSilva:2019ykt,Cavalcanti:2020obq}. Notice that the standard Dirac adjoint is recovered by imposing $\Delta = \mathbbm{1}$. However, guided by the discussions around Dirac spinors in Ref. \cite{BuenoRogerio:2019kgd}, more specifically for spinors belonging to class-2, which stand for eigenspinors of parity operator, the right way to define the adjoint structure is observing the relation $\Delta \psi = P\psi = \pm\psi$. 

Nonetheless, as already exhaustively verified, when a new adjoint structure is imposed, the new elements of the Clifford algebra reads \cite{Vaz:2016qyw,HoffdaSilva:2016ffx}
\begin{equation}\label{newcliffordbasis}
\tilde{\Gamma} = \lbrace \mathbbm{1}, \gamma_{\mu}, \Delta i\gamma_{\mu}\gamma_{\nu}\Delta, \Delta\gamma_5\gamma_{\mu}\Delta, \Delta\gamma_{0123}\Delta\rbrace, 
\end{equation}
and thus, consequently, a new set of bilinear forms are defined

\begin{subequations}
\label{newbilinear}
\begin{align}
\tilde{\sigma}=&\stackrel{\neg}{\psi}\psi \in\Omega^0(M),  \\
\tilde{\mathbf{J}}=&\stackrel{\neg}{\psi}\gamma_{\mu}\psi\;\gamma^{\mu} \in\Omega^1(M),\\
\tilde{\mathbf{S}}=&\stackrel{\neg}{\psi} \Delta i\gamma_{\mu}\gamma_{\nu}\Delta\psi\; \Delta\gamma^{\mu}\wedge\gamma^{\nu}\Delta \in\Omega^2(M),\\ 
\tilde{\mathbf{K}}=&\stackrel{\neg}{\psi}\Delta\gamma_{\mu}i\gamma_{0123}\Delta\psi\;\gamma^{\mu} \in\Omega^3(M),\\
\tilde{\omega}=&-\stackrel{\neg}{\psi}\Delta\gamma_{0123}\Delta\psi \in\Omega^4(M). 
\end{align} 
 \end{subequations}

Notice that the bilinear forms $\tilde{\omega}$, $\tilde{\mathbf{K}}$ and $\tilde{\mathbf{S}}$ has a new contrasting structure when compared with \eqref{covariantes}. The above structures lead accordingly to the bilinear densities $\tilde{\rho}_{i} =\stackrel{\neg}{\psi}\tilde{\Gamma}_i\psi$. One should also note that such slight modifications on the bilinear structures guarantee that the FPK identities \eqref{fpk1} and \eqref{fpk2} are satisfied. A hint towards the physical interpretation for the bilinear structures given in Eq.\eqref{newbilinear} can be seen in \cite{CoronadoVillalobos:2020yvr}. However, such new introduced dual forces to re-examine Eqs. \eqref{Z} and \eqref{3}, in the lights of the new adjoint structure. For the dual structure introduced in \eqref{3dual} the counterpart of \eqref{novodual} is the give by  
\begin{eqnarray}
\stackrel{\neg}{\psi} = \frac{1}{4N^*}\xi^{\dag}\mathbf{Z}^{\dag}\Delta^{\dag}\gamma_0 e^{i\theta},
\end{eqnarray}
while the counterpart of \eqref{bilinearreconstruido} may be displayed in the following fashion
\begin{equation}
\tilde{\rho}_i=  \frac{1}{16|N|^2}\xi^{\dag}\mathbf{Z}^{\dag}\Delta^{\dag}\gamma_0 \tilde{\Gamma}\mathbf{Z}\xi.
\end{equation}
% the general forms above has as an particular case when $\Delta=\mathbbm{1}$, please, check \eqref{3dual}.
The dual structure modification performed above leads to the new (deformed) Fierz aggregate ($\tilde{\mathbf{Z}}$), given by
\begin{eqnarray}
\tilde{\mathbf{Z}}=\sigma + \mathbf{J} +i\tilde{\mathbf{S}}+i\tilde{\mathbf{K}}\Delta\gamma_{0123}\Delta+\tilde{\omega} \Delta\gamma_{0123}\Delta \label{Znew}
\end{eqnarray}
and the FPK identities must be re-written in terms of $\tilde{\mathbf{Z}}$, yielding
\begin{align}\label{fpk2new}
\tilde{\mathbf{Z}}^{2}=4\sigma \tilde{\mathbf{Z}},\nonumber\\
\tilde{\mathbf{Z}}\gamma_{\mu}\tilde{\mathbf{Z}}=4J_{\mu}\tilde{\mathbf{Z}},\nonumber\\
\tilde{\mathbf{Z}}\Delta i\gamma_{\mu\nu}\Delta\tilde{\mathbf{Z}}=4S_{\mu\nu}\tilde{\mathbf{Z}},\\
\tilde{\mathbf{Z}}i\Delta\gamma_{0123}\Delta\gamma_{\mu}\tilde{\mathbf{Z}}=4K_{\mu}\tilde{\mathbf{Z}},\nonumber\\
\tilde{\mathbf{Z}}\Delta\gamma_{0123}\Delta\tilde{\mathbf{Z}}=-4\omega \tilde{\mathbf{Z}}.\nonumber
\end{align}

Analyzing the inversion theorem and the FPK identities for duals defined upon $\mathcal{P}$, $\mathcal{C}$, $\mathcal{T}$ and its combinations, it furnishes the results displayed in Table \ref{inversion_table1}. We highlight that those results show that the inversion theorem is only valid for some specific cases, depending on the dual structure. Besides, the most relevant symmetry analyzed here is the $\mathcal{P}$ symmetry, accordingly discussions about duals in \cite{HoffdaSilva:2022ixq}, and new duals proposed in \cite{Ahluwalia:2019etz,Ahluwalia:2022ttu,Ahluwalia:2020jkw,Ahluwalia:2020miz}. Finally, it is worth mentioning that the results hold, in general, for any arbitrary spinor, satisfying only the constraint of the spinorial components being eigenstates of the helicity operator, with no other requirement on the $(1/2, 0)$ and $(0, 1/2)$ representation spaces being necessary.

\begin{table}[!h]
    \centering
    \begin{tabular}{c|ccccccccc}
&&\multicolumn{2}{c}{\textbf{Inversion Theorem}}&&\multicolumn{2}{c}{\textbf{FPK identities}}&&\multicolumn{2}{c}{\bf{$\rho_{i}\in \mathbb{R}$}}\\
        Operator && Regular & Singular &  & Regular & Singular &  & Regular & Singular \\ \midrule
        $\Delta=\mathcal{P}$ && $\checkmark$ & $\checkmark$ &  & $\checkmark$ & $\checkmark$ &  & $\checkmark$ & $\times$ \\ 
        $\Delta=\mathcal{C}$ && $\times$ & $\times$ &  & $\times$ & $\times$ &  & $\times$ & $\checkmark$ \\ 
        $\Delta=\mathcal{T}$ && $\checkmark$ & $\checkmark$ &  & $\times$ & $\times$ &  & $\times$ & $\times$ \\ 
        $\Delta=\mathcal{PC}$ && $\times$ & $\times$ &  & $\times$ & $\times$ &  & $\times$ & $\times$ \\ 
        $\Delta=\mathcal{PT}$ && $\times$ & $\times$ &  & $\times$ & $\times$ &  & $\times$ & $\times$ \\ 
        $\Delta=\mathcal{CT}$ && $\times$ & $\checkmark$ &  & $\times$ & $\checkmark$ &  & $\times$ & $\times$ \\ 
        $\Delta=\mathcal{CPT}$ && $\times$ & $\times$ &  & $\times$ & $\times$ &  & $\times$ & $\times$ \\ \bottomrule
    \end{tabular}
    \caption{Validity of Inversion theorem and FPK identities for different duals.}
    \label{inversion_table1}
\end{table}

We shall make some comments concerning the symmetry operator's essential aspects. Firstly, we note that for the standard case ($\Delta=\mathbbm{1}$) all the constraints are satisfied, as expected. In table \ref{inversion_table1}, the $\Delta=\mathcal{CT}$ case for singular spinors is potentially interesting, while the $\Delta=\mathcal{P}$ case is the most relevant one. For spinors not belonging to the eigenstates of $\mathcal{P}$, the unique caveat is the reality of bilinears to the singular spinor type. More specifically, $\mathbf{J}$ is the odd bilinear. This caveat, however, is not problematic: when a given spinor is an eigenspinor of $\mathcal{P}$, it satisfies the Dirac equation, and, as a consequence, $\mathbf{J}$ is the conserved current. This is not the case for spinors that are not eigenspinors of $\mathcal{P}$; therefore, its conserved current is not extracted from the Dirac equation or Lagrangian. This analysis may be applied to the $\mathcal{C}$ symmetry to some extent. In this last case, $\mathbf{J}$ is only real for spinors that are eigenstates of $\mathcal{C}$, only a particular class of singular spinors.

\section{Extensions of the Lounesto classification} \label{extension_lounesto}

There is a program extending the Lounesto classification by relaxing the role played by $\stackrel{\sim}{\mathbf{J}}$. When the spinor field does not obey the Dirac equation, $\stackrel{\sim}{\mathbf{J}}$ is not the associated conserved current, as mentioned. This program started in \cite{CoronadoVillalobos:2015mns}. The crucial point is that the FPK identities are satisfied, and the inversion theorem is valid for these new classes. They are mathematically sounding, thus potentially hiding the description of new physics. Here we present a more comprehensive picture by showing two new possible classes according to the previous criteria. The results are shown in table \ref{new_classes}. Once new mathematical classes are established, it is only natural to ask whether they are physically realized. For instance, in Ref. \cite{Ahluwalia:2020miz}, a concrete example of a spinor populating class 10 (though reducible) is presented. 

\begin{table}[h]
\centering
\begin{tabular}{c|cccccccccccc}
% \toprule
Bilinears & Class 1 & Class 2 & Class 3 & Class 4 & Class 5 & Class 6 & Class 7 & Class 8 & Class 9 & Class 10 & Class 11 \\
\midrule
$\stackrel{\sim}{\sigma}$ & $\neq 0$ & $\neq 0$ & 0 & 0 & 0 & 0 & 0 & 0 & 0 & $\neq 0$ & $\neq 0$ \\
$\stackrel{\sim}{\mathbf{J}}$ & $\neq 0$ & $\neq 0$ & $\neq 0$ & $\neq 0$ & $\neq 0$ & $\neq 0$ & 0 & 0 & 0 & 0 & 0 \\
$\stackrel{\sim}{\mathbf{S}}$ & $\neq 0$ & $\neq 0$ & $\neq 0$ & $\neq 0$ & $\neq 0$ & 0 & $\neq 0$ & $\neq 0$ & 0 & $\neq 0$ & 0 \\
$\stackrel{\sim}{\mathbf{K}}$ & $\neq 0$ & $\neq 0$ & $\neq 0$ & $\neq 0$ & 0 & $\neq 0$ & $\neq 0$ & 0 & $\neq 0$ & 0 & 0 \\
$\stackrel{\sim}{\omega}$ & $\neq 0$ & 0 & $\neq 0$ & 0 & 0 & 0 & 0 & 0 & 0 & $\neq 0$ & $\neq 0$ \\
\bottomrule
\end{tabular}
\caption{Non-zero values of bilinear forms by class.}
\label{new_classes}
\end{table}

More generally, the dual structure presented as in \eqref{novodual}, while preserving FPK identities and the inversion theorem, helps to access some of the new classes. It is remarkable that classes 7 and 8 are populated by a dual structure different from the usual. The results are depicted in table \ref{operator_classes}. It is worth mentioning once again that $\stackrel{\sim}{\mathbf{J}}=0$ imposes a reinterpretation of the bilinears for the new classes, an exciting issue that is still open. As seen, spinors in classes 9 and 11 are not reachable for any theory built upon whichever combination of discrete symmetries. Of course, another dual structure may reach these two elusive classes.

\begin{table}[H]
    \centering
    \begin{tabular}{c|cccccccc}
       Operator  && Class 7 & Class 8 & Class 9 & Class 10  & Class 11&& \\ \bottomrule%\midrule
        $\Delta=\mathcal{P}$ &&  $\times$ &  $\times$ &  $\times$ & $\checkmark$ &  $\times$ &      \\ 
        $\Delta=\mathcal{C}$ &&  $\times$ & $\checkmark$ &  $\times$ &  $\times$ &  $\times$ &      \\ 
        $\Delta=\mathcal{T}$ && $\checkmark$ & $\checkmark$ &  $\times$ &  $\times$ &  $\times$ &    \\ 
        $\Delta=\mathcal{PC}$ && $\checkmark$ &  $\times$ &  $\times$ &  $\times$ &  $\times$ &     \\ 
        $\Delta=\mathcal{PT}$ && $\checkmark$ &  $\times$ &  $\times$ &  $\times$ &  $\times$ &      \\ 
        $\Delta=\mathcal{CT}$ &&  $\times$ &  $\times$ &  $\times$ &  $\times$ &  $\times$ &     \\ 
        $\Delta=\mathcal{CPT}$ &&  $\times$ &  $\times$ &  $\times$ & $\checkmark$ &  $\times$ &    \\ \bottomrule
    \end{tabular}
    \caption{New classes accessible for combinations of the discrete operators.}
    \label{operator_classes}
\end{table}

\section{Concluding Remarks and outlooks}\label{remarks}

The intricate mathematical structure underlying spinor fields have been an active research subject for almost a century since Cartan's seminal work \cite{cartan2012theory}. In this regard, the present paper contributes to understanding one of the fundamental and little exploited properties of spinors in field theories, namely the alternative spinorial adjoint. Specifically, we have explored the validity of the inversion theorem and the FKP identities for a class of duals by applying the discrete symmetry operators, $\mathcal{C}, \mathcal{P}, \mathcal{T}$, and their combinations. By shedding new light on the role of discrete symmetries and their implications for duals, we have extended the classes of the Lounesto classification, uncovering new possibilities of investigations by finding concrete realizations of representatives of the new classes.   

Our findings contribute to the ongoing discussion surrounding modern spinor field theories central to contemporary physics. In addition, the new duals we have introduced have potential applications in high-energy physics, particularly beyond the standard model. Expanding the Lounesto classification could help understand those alternative theories better, providing a solid and physically appealing classification scheme.

\section{Acknowledgements}

R.T.C. thanks Unesp—AGRUP for the financial support. J.M.H.d.S. thanks CNPq (grant No. 307641/2022-8) for the financial support.

\bibliography{Inversion_Theorem_biblio}

\end{document}